\documentclass[12pt,a4paper]{article}
\usepackage[hyperref]{acl2018}
\usepackage{times}
\usepackage{latexsym}
\usepackage{url}
\usepackage{graphicx}
\usepackage{amsmath}
\usepackage[skip=9pt]{caption}
\DeclareCaptionFormat{myformat}{\fontsize{9}{9}\selectfont#1#2#3}
\usepackage{refstyle}
\usepackage{multirow}
\usepackage{caption}
\usepackage{longtable}

\aclfinalcopy 

\title{A Neural Network Based Explainable Recommender System}

 \author{Jionghao Lin\\
 Monash University
  \\
  Clayton, VIC, Australia
    \\
 jlin0028@student.monash.edu
  \\
 \\\And
  Yiren Liu \\
  University of Electronic  \\Science and Technology of China
  \\
  Chengdu, Sichuan, China
    \\
 yirenliu@std.uestc.edu.cn
  \\
 \\}

\begin{document}
\maketitle

\begin{abstract}
Recommendation system could help the companies to persuade users to visit or consume at a particular place, which was based on many traditional methods such as the set of collaborative filtering algorithms. Most research discusses the model design or feature engineering methods to minimize the root mean square error (RMSE) of rating prediction, but lacks exploring the ways to generate the reasons for recommendations. This paper proposed an integrated neural network based model which integrates rating scores prediction and explainable words generation. Based on the experimental results, this model presented lower RMSE compared with traditional methods, and generate the explanation of recommendation to convince customers to visit the recommended place.

\end{abstract}
\vskip 0.2in
\section{Introduction}

User-generated content such as rating and reviews about a place is gradually becoming a key factor for other people making a decision. Some platforms such as Yelp utilize recommender system to analyze the user preferences based on the reviews and rating to recommend a new place for other customers. According to \cite{liu_experimental_2017}, they introduce a phrase point-of-interests (POIs) to represent a set of places such as restaurant and tourist attractions where users are interested. In this paper, we will continue to use this phrase and focus on restaurant rating prediction and explanation generation. Currently, much empirical research focuses on model design or feature engineering methods to improve model performance rather than discussing the ways to generate recommended reasons.  This project proposed an integrated neural network based explainable recommender system, instead of using traditional methods such as user-based models, and Matrix Factorization, to generate explainable opinion-aspect pairs for a user, helping them explore their unfamiliar places, and predict rating score from that user for each POI and recommend POI with high predicted rating to each user. Our model could generate user latent preference from embedding layer in neural network after training rating predictions for POIs Instead of training models on user-POI rating matrix directly. The main contributions of this paper are summarized as follows:
\begin{itemize}

\item Predict rating stars for POIs based on user-POI rating matrix using neural-based model.
\item Generate opinion-aspect pairs for users to explain reasons for recommended POI.
\item Compare the model performance with the classical recommender system methods.
\item Propose a way to evaluate the prediction of explainable pairs.
\end{itemize}

\noindent
In the rest of the paper, we will discuss the related work in Section 2. Provide the details of data preprocessing, feature extraction, and data exploration in section 3. We will introduce the proposed model and other traditional models in Section 4. In Section 5 and Section 6, We discuss the method to evaluate the RMSE of predicting rating stars and the f-score of predicting explainable tuples. Then, we present the results including comparison with other baseline models. Finally, we conclude findings, limitation, and future work of this paper in section 7.

\section{Related Work}
Much empirical research focuses on either rating stars treating the scores as numerical data or explainable text generation offering the features of the POIs, and those studies adopt different models, features and evaluation methods \cite{liu_experimental_2017,bao_location-based_2012,zhang_orec:_2015,liu_personalized_2013,duan_poi_2017,rendle_bpr:_2009}. This project is motivated by these empirical studies on rating stars prediction, and we proposed an integrated model which can predict rating stars and generate explainable opinion-aspect pairs for users.
\subsection{Collaborative Filtering for Rating Prediction}
\vskip 0.05in
\noindent
Collaborative Filtering is commonly used in recommender system\cite{chen_context-aware_2014,konstas_social_2009,su_survey_2009,wang_collaborative_2011} --- exploiting similarity among the preference of users to generate recommendations. Research is done using users’ historical data to predict ratings, and deliver the results of recommendation to users \cite{balabanovic_fab:_1997,rennie_fast_2005}. The traditional methods in rating prediction are the user-based model, and matrix factorization(MF). According to \cite{ricci_recommender_2015}, they introduced the user-based method which predicts the rating scores based on the other users who have similar preferences. Regarding matrix factorization, Ricci  et  al. mentioned the applications of two commonly used methods: Singular Value Decomposition(SVD) and Non-negative Matrix Factorization(NMF) \cite{ricci_recommender_2015}. Additionally, many researchers utilized neural-based collaborative filtering model for rating prediction \cite{van_den_oord_deep_2013,wang_collaborative_2015}. However, there are many limitations for the traditional algorithms such as scalability problems, and lacking bias terms. As for neural collaborative filtering, it performs better on rating prediction, but there is no description of generating explainable text in empirical works. 

\subsection{Explainable Recommendation System}
\vskip 0.05in
\noindent
Explainable recommendation system (ERS) gradually attracts more researchers to explore the ways to generate convincing explanations for users. According to \cite{hou_explainable_2018}, they state two forms of ERS to output explanations. One is to extract noun words or phrases to represent item features, and there are many works utilizes this form for explainable recommendation. However, Hou et al. claim that it lacks consideration of sentiment for each aspect summarized from other visited customers which is less convincing for future users. Another method is to generate a set of words containing opinion and aspect phrases from reviews to explain fine-grained aspect information for future users \cite{hou_explainable_2018}. In this project, we investigated the second form of ERS, and summarized many empirical works about this type as follows:
\vskip 0.05in
\noindent
Baccianella et al. introduced a method using SentiWordNet in extracting opinion words and quantifying the sentiment polarity and strength of each word to the POI \cite{baccianella_sentiwordnet_2010}. Pero et al. proposed a recommendation system by integrating opinion information with the rating score for the POI \cite{pero_opinion-driven_2013}. Chen et al. demonstrated an algorithm about tensor decomposition to predict rating based on opinion-aspect pairs in review over multiple domains \cite{chen_context-aware_2014}. However, these method lacks of fine-grained sentiment differences based on some specific aspect of POIs \cite{hou_explainable_2018}.

\vskip 0.05in
\noindent
Though given plenty of research focusing on POI rating prediction in recommender systems, less research has investigated the recommender system offering the fine-grained explanation for specific aspects of POIs. Former Collaborative Filtering research focuses on improving the precision of recommendation but lacks information to help target users better understand the proposed recommendation. Our proposed model could resolve the mentioned limitations from empirical works, which integrate rating prediction and explainable pairs generation, and provide better performance compared with other traditional models.

\section{Data}
\subsection{Data Preprocessing and Feature Extraction}
The data is composed of business ID, user ID, rating score (range from 1 to 5), reviews, and date in the city Pittsburgh from Yelp \cite{Yelp_yelpDataset_2018} challenge dataset. Considering the impact of time,we set 2017 (full year) as the training dataset and 2018 (until June) as test dataset which contains 26918 reviews (from 13225 users to 450 restaurants).  After tokenizing and POS tagging, review texts are transformed into syntax relations (shown in Figure 1) by utilizing the spaCy CNN dependency parsing model \cite{kiperwasser_simple_2016,goldberg_dynamic_2012}, and these relation pairs containing user’s opinions towards aspects are extracted in the form of tuples liked\textless opinion, aspect\textgreater. 
\vskip 0.05in
\noindent
Additionally, SentiWordNet \cite{baccianella_sentiwordnet_2010} is incorporated to quantify the opinion sentiment polarity and strength, which will be described in detail in the evaluation section.
\captionsetup{format=myformat}
\setlength\belowcaptionskip{-3ex}
\begin{figure}[h!]
\centering
\includegraphics[width=0.45\textwidth]{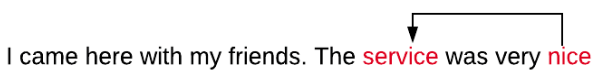}
\caption{Aspect-opinion Pairs Examples} 
\label{fig:opinion}
\end{figure}

\vskip 0.1in
\noindent
Finally, after lemmatizing and lowercasing the words, we use 100-dimension Glove word embedding pretrained by twitter corpus \cite{pennington_glove:_2014} , and then horizontally concatenate the word embedding to tuple embedding. 

\subsection{Data Exploration Analysis}
After preprocessing the data, the stars distribution is visualized in Figure 2 which indicates that users rarely give 1 or 2 stars for POIs.

\captionsetup{format=myformat}
\setlength\belowcaptionskip{-3ex}
\begin{figure}[h!]
\centering
\includegraphics[width=0.45\textwidth]{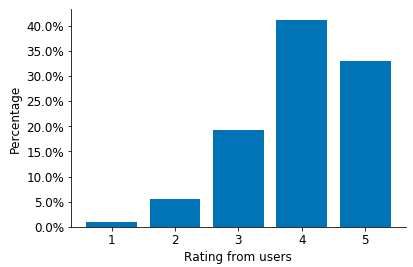}
\caption{Rating Distribution Bar Chart} 
\label{fig:opinion}
\end{figure}

\vskip 0.05in
\noindent
Then, we explore the opinion-aspect pairs distribution in the space of frequency and average rating. In Figure 3, these pairs are visualized based on the frequency greater than 700 and average stars that they appeared.

\captionsetup{format=myformat}
\setlength\belowcaptionskip{-3ex}
\begin{figure}[h!]
\centering
\includegraphics[width=0.5\textwidth]{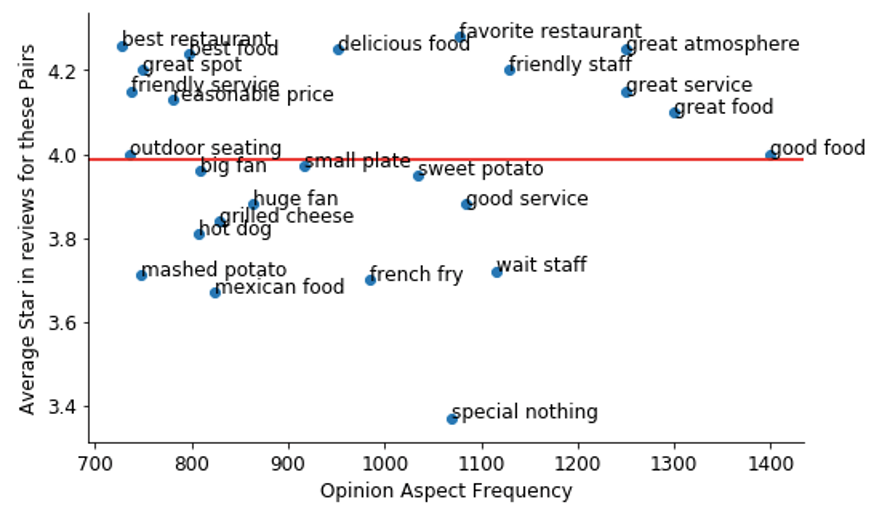}
\caption{Word Pairs} 
\label{fig:opinion}
\end{figure}

\vskip 0.05in
\noindent
Additionally, there are many examples having comparable high average stars which accord with their opinion words such as 'best restaurant', and 'best food'. On the other hand, many tuples such as 'good service' have positive opinion but present the position lower than the average stars line. This case will be discussed in our limitation part.

\section{Method}
There are many traditional model framework used for rating prediction. In this section, we experimented with four algorithms: neural-based model, K nearest neighbor, SVD matrix factorization and NMF matrix factorization to predict the rating stars that a user would give for a POI based on the ratings of reviews posted in the past. Then, we demonstrate the ways of our models to generate opinion-aspect pairs for target users after completing the model training for RMSE.

\subsection{Baseline Model for Rating Prediction}
\subsubsection{User-based collaborative filtering}
\vskip 0.05in
The first baseline model is a user-based collaborative filtering algorithm based on the K nearest neighbor (KNN) algorithm. According to \cite{ricci_recommender_2015}, they claim that this method utilizes a ratings matrix to compute centered cosine similarity (Pearson relationship) between each user. Then, predict the rating score for the POI where the target user has never been there before based on the weight of similarity and scores from other top K users who are most similar to target user \cite{ricci_recommender_2015}. This method could provide decent performance in rating prediction. However, this method is based on the KNN algorithm which will suffer serious scalability problems when the number of users and the number of items increase \cite{shani_evaluating_2011,ricci_recommender_2015}.

\subsubsection{Matrix Factorization}
Matrix factorization is a commonly used collaborative filtering method \cite{koren_factor_2010} which decomposes the user-restaurant rating matrix into two latent feature matrices of user and restaurant and can be used to predict the rating stars of restaurants the users who have not visited. Singular value decomposition (SVD) and non-negative matrix factorization (NMF) are two commonly used matrix factorization methods and will be used as baseline models in this paper. 
\vskip 0.05in
\noindent
Matrix Factorization algorithms are known to have high computational efficiency and generate user and item embedding containing information about users’ preferences and item features, which can be extracted for further analysis. However, traditional Matrix Factorization methods have a hindered performance compared to neural based models due to their ignorance of users’ and items’ bias terms.

\subsection{Integrated Neural Collaborative Filtering}
Our proposed neural-based model is motivated by \cite{he_neural_2017}, which is comprised of two parts --- rating prediction and explainable pair generation (shown in Figure 4). As for rating prediction, the model is trained using the user-POI rating matrix and predicts the rating stars of each user on each POI. Concerning explainable pairs generation, we use the user embedding learned from embedding layer to find the top-K users who have the most similar preferences to the target user and then generate explanations based on their reviews over each specific POI. Our model resolves the issues mentioned in the baseline models and considers the bias terms when calculating users’ and POIs’ embedding. Additionally, the most similar users could be detected from user embedding instead of using other formulas to compute the correlation.
\captionsetup{format=myformat}
\setlength\belowcaptionskip{-3ex}
\begin{figure}[h!]
\centering
\includegraphics[width=0.5\textwidth]{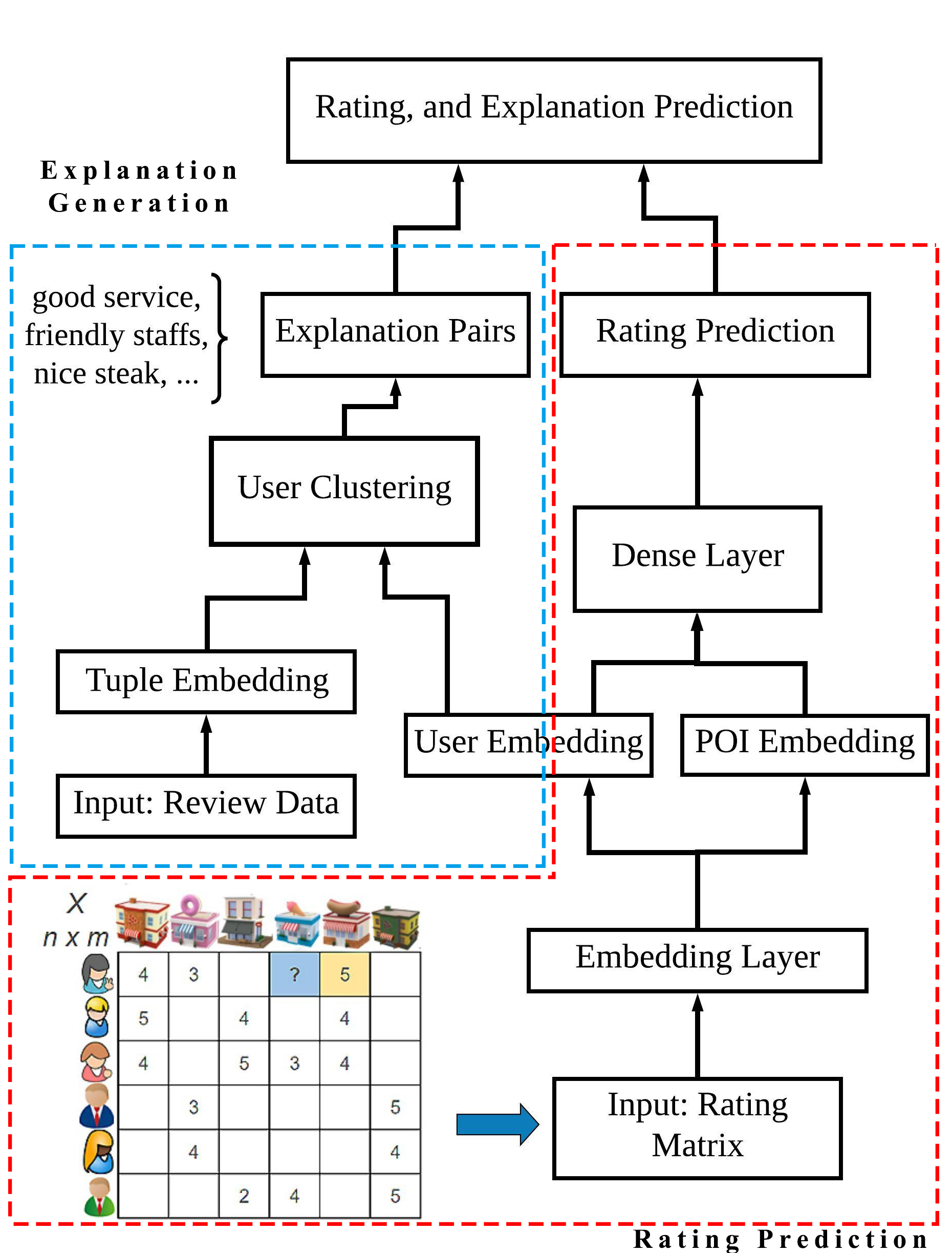}
\caption{Integrated Neural Collaborative Filtering} 
\label{fig:opinion}
\end{figure}

\subsubsection{Rating Prediction}
We predict ratings for POIs that a user has not visited before. In rating prediction part (shown in Figure 4. right block), our model takes user-POI matrix as the input and predicts rating stars for the user over the POI. After training, the model vectorizes the latent preferences of users about POIs in the embedding layers \cite{he_neural_2017}, which will be used as the input for the explanation generation. (shown in Figure 4. left block).
\vskip 0.05in
\noindent
After the score of each POI is predicted, the top-K POIs with highest ratings that the user has not visited in the past are offered as the recommended POIs for the user.
Our neural-based model is established using Keras in Python. The model is trained using the review data of restaurants in Pittsburgh in 2017 and validated using data in 2018 and sets RMSE as the loss. We trained our model in 200 epochs, converging at around 0.17 (shown in Figure 5).

\captionsetup{format=myformat}
\setlength\belowcaptionskip{-3ex}
\begin{figure}[h!]
\centering
\includegraphics[width=0.5\textwidth]{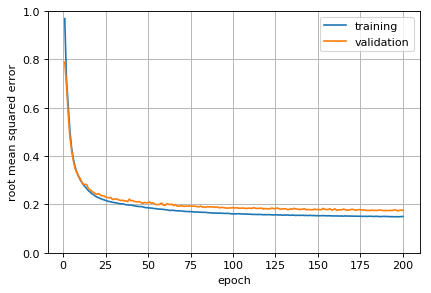}
\caption{Learning curve for the NCF method} 
\label{fig:opinion}
\end{figure}

\subsubsection{Explainable Pairs Generation}
As the neural model completed the training phase, the user embedding layer is extracted from user-POI matrix which contains the information about the latent preferences of each user. After determining the POIs which will be recommended to each target user, we use the extracted user embedding to find the top-K most similar users of our target user who have been to the same POIs. The similarity between users is measured by calculating the cosine similarity using the user embedding. Then, we extract the pairs from the reviews of the top-K users generated for the POIs using dependency parsing and sample from these pairs as the explanation provided to the target user. 

\section{Model Evaluation}
\subsection{Evaluation of Rating Prediction}
The introduced recommendation strategy relies on the prediction of ratings over unvisited POIs. To evaluate the rating prediction model, we use root mean square error (RMSE) to measure its performance \cite{lu_recommender_2015}. As mentioned in the Section 3, we treat 2017 as training set and 2018 as testing set, so we predict the rating stars that a user will give for an unvisited POI in 2018, and use the ratings in test dataset to measure RMSE of the result.
 \vskip 0.1in
\begin{equation}
RMSE=\sqrt{\frac{\Sigma_{i=1}^{n}(P_{i}-R_{i})^2}{n}}
 \end{equation}
 \vskip 0.2in
\subsection{Evaluation of Explanation Generation}
To evaluate the quality of the generated explanations,  we calculate the cosine similarity between the pairs that we predicted for POIs and pairs extracted from the user review of corresponding POIs in the test set. Since many opinion-aspect pairs present the same meaning but different word, this project sets the cosine similarity greater than 0.8 as true, otherwise false. Then, we calculate the median number of pairs of the target user  in the training dataset and predict same amount of pairs according the median number.
\vskip 0.05in
\noindent
The way that we propose for evaluation of explanation prediction will bring some problems including false positive and false negative. However, if we use exact match, it will ignore the same meaning between tuples with semantic similarity but different words such as 'great beef' and 'great steak'. Since we focus more on related pairs recommendation to users, we decide to propose this way to evaluate explainable pairs. After examining many tuples in the dataset, we found that setting the cosine similarity threshold as 0.8 can generally distinguish relevant pairs from irrelevant ones (shown in Table 1).

\begin{table}[h!]
\small
\begin{tabular}{p{2.34cm}|p{2.15cm}|p{1.2cm}|p{0.55cm}}
\hline
Pair 1 & Pair 2 & Similarity&Label\\
\hline\hline
\rule{0pt}{15pt}('good', 'food') & ('nice', 'food') & 0.969 (+)&True\\
\hline
\rule{0pt}{15pt}('great', 'sides') & ('great','snacks') & 0.821 (+)&True\\
\hline
\rule{0pt}{15pt}('great', 'beef') & ('nice', 'steak') & 0.855 (+)&True\\
\hline
\rule{0pt}{15pt}('bad', 'food') & ('bad', 'pricing') & 0.761 (+) & False\\
\hline
\rule{0pt}{15pt}('good', 'service') & ('bad', 'service') & 0.952 (-) & False\\
\hline
\rule{0pt}{15pt}('friendly', 'staff') & ('cool', 'vibe') & 0.579 (+) & False\\
\hline
\end{tabular}
\caption{Examples for Evaluating Tuples}
\label{Table:1}
\end{table}
\vskip 0.1in
\noindent
Furthermore, we found that many pairs have the same aspect but different opinion words like 'good' or 'bad' which cannot be effectively distinguished simply using GloVe vectors. Therefore, this paper introduces a cosine similarity based F-score with sentiment penalty and uses it as the evaluation method. In this method, if two words have similarity greater than 0.8 but with opposite sentiment polarity, we consider this as a false case (shown in equation 2). In Table 1, the cosine similarity with (+) means two words have the same sentiment polarity and (-) means different. Although 'good service' and 'bad service' have quite high similarity (0.952), this case will be labelled with False because the sentiment penalty presents (-). We will discuss more details about the impact without using sentiment penalty in Section 6.
\vskip 0.05in
\noindent
To calculate the F-score during evaluation, we first determine the median number of pair counts of each user generated from reviews for each POI in training set. We set the median count number as the number of explanation pairs generated using our model individually for each user. According to \cite{shani_evaluating_2011}, they listed ways to calculate the recall rate (shown in equation 3), precision (shown in equation 4), and F-score (shown in equation 5). Then, we following these ways to calculate F-score for our case.
\vskip 0.05in
\noindent
The evaluation method accommodates the error caused by using GloVe vectors during explanation evaluation. The modified F-score consolidates sentiment penalty requiring a match with a cosine similarity under 0.8 and both opinion words having the same sentiment polarity, which can be written as follows:

\begin{equation}
  \small
n_{i}=\begin{cases}
			1, & \text{if $s_{i}$}\cdot\overline{s_{i}} > 0 \text{ and } similarity > 0.8\\
            0, & \text{otherwise}
		 \end{cases}
 \end{equation}

  \begin{equation}
  \small
recall_{n_{i}}=\frac{number\ of\ True\ predicted\ pairs}{total\ number\ of\ pairs\ in\ reviews}
 \end{equation}
 \vskip 0.1in
 \noindent
  \begin{equation}
  \small
precision_{n_{i}}=\frac{number\ of\ True\ predicted\ pairs}{total\ number\ of\ pairs\ predicted}
 \end{equation}
 \vskip 0.1in
 \noindent

 \begin{equation}
   \small
F_{1}=\frac{2\cdot recall_{n_{i}}\cdot precision_{n_{i}}}{recall_{n_{i}}+ precision_{n_{i}}}
 \end{equation}
 \vskip 0.1in
 \noindent
Where $s_{i}$ is the sentiment score of opinion in the pair generated by our model predicted using SentiWordNet \cite{baccianella_sentiwordnet_2010}, $\overline{s_{i}}$ is the  sentiment score of opinion in the pair extracted from user’s review.

\section{Results Discussion and Analysis}
We apply the same dataset to the three baseline methods: KNN, SVD matrix factorization, and NMF matrix factorization, which refer to the algorithms in \cite{ricci_recommender_2015,luo_efficient_2014}. The result presents that our proposed integrated neural collaborative filtering method (NCF) outputted the lowest RMSE compared with other methods (shown in Table 2).
\vskip 0.1in
\begin{table}[h!]
\centering
\begin{tabular}{ ||c|c|c|c|| } 
\hline
NCF & KNN & MF - SVD & MF - NMF\\
\hline
 0.1733 & 0.2555 & 0.3032 & 0.3890 \\ 
\hline
\end{tabular}
\caption{Rating Prediction Results}
\label{Table:2}
\end{table}
\vskip 0.1in
\noindent
Then, we evaluate our model in terms of explainable pairs over this dataset. The baseline model we introduced is to randomly sample pairs extracted from all users' reviews under the predicted POIs which correspondes to human random guess. The random sampling method provides information extracted from historical reviews of the recommended POI but does not consider users' preferences. By using this as a baseline, we can validate how much improvement our model achieved through incorporating information of users' preferences. To collect unbiased result, we use ten folds cross-validation for each experiment, and the result shows (in Table 3) that NCF with sentiment penalty outcomes an F-score of 0.5088, which is higher than 0.0349 of random sampling. As we mentioned the sentiment penalty in the process of evaluating our explainable results before, it could help to modify the labeling results. Since 'bad service' and 'good service' have 0.952 similarity but they present different sentiment, we add sentiment penalty to label these cases as False. Therefore,  after incorporating the sentiment penalty, the F-score drops from 0.5696 to 0.5088 (shown in Table 3), which indicates the deficiency without applying penalty.

\begin{table}[h!]
\centering
\begin{tabular}{ ||p{2cm}|p{2cm}|p{2cm}|| } 
\hline
NCF & NCF* & Baseline\\ 
\hline
 0.5696 & 0.5088 & 0.0349 \\ 
\hline
\end{tabular}
      \begin{flushleft}
      \small
      * represents using sentiment penalty
      \end{flushleft}
\caption{Explanation Generation Results}
\label{Table:3}
\end{table}

\vskip 0.1in

\section{Conclusion}
This paper introduces a neural-based explainable recommender system comprised of 2 parts: providing recommendation to users based on rating prediction and integrating the recommendation with explanations that can convince users to visit the proposed recommendation. We deployed a neural collaborative filtering model to predict ratings for the target user. Further, we utilize the user embedding layer which contains information on users' latent preference to generate explanations based on users' historical review data. Then, the neural based rating prediction model is evaluated using Yelp dataset and compared with commonly used collaborative filtering methods. Finally, we introduce cosine similarity based F-score with sentiment penalty to evaluate our explanation generation method. Result shows that our model achieves a better performance in rating prediction than classical collaborative filtering model and a significant improvement over the random sampling baseline in explanation generation.
\vskip 0.05in
\noindent
\textbf{Limitation and Future Work}
\vskip 0.1in
\noindent
The evaluation results of rating and explanation present that our model is useful to solve some problems and could be generalized to other types of POI recommendation such as hotels and tourist attractions. However, there are also some limitations which are discovered during our experiments. We listed these limitations and discussed the ways to improve our model performance in future work.
\vskip 0.05in
\noindent
In opinion-aspects pair extraction, we only explore the aspect-opinion pairs in the form of adjectival modifier and noun using dependency parsing. Current method incorporated excludes part of POI features that are not presented in the form of opinion-aspect pairs, such as 'cafe near the river' and 'restaurant in the city center'. If we consider higher level n-gram (n\textgreater2) for extracting opinion-aspect pairs by using dependency parsing method, more noise pairs will also be extracted. Hence, we consider using deep learning model for explanation generation in the future.
\vskip 0.05in
\noindent
The rating prediction phase only considers the rating scores between users and POIs. Since the customer reviews also relate to the rating stars, we will incorporate other information about users such as social links to improve the model performance in future work. Review information may be incorporated by adding another embedding layer which encodes users' reviews and import reviews as input of the layer.
\vskip 0.05in
\noindent
Our model assumes that the users have visiting records before, and continue to use the website which contains user generated content about POIs. Therefore, we will explore the methods for the users who don't have any visited record in browser or website.

\vskip 0.1in
\noindent
\textbf{Contribution}
\vskip 0.1in
\noindent
The authors of this project are Jionghao Lin and Yiren Liu. Both of us contribute to idea brainstorming, code implementation, and report writing. Jionghao was mainly responsible for preprocessing dataset, model evaluation and generate graphs/figures for the report. Yiren Liu mainly focused on model designing and implementation.



\bibliographystyle{acl_natbib}
\bibliography{ref.bib}

\end{document}